\documentclass[twocolumn]{aastex63}
\usepackage{verbatim}
\usepackage{comment} 

\def\be{\begin{equation}}
\def\ee{\end{equation}}

\def\gsim{\lower.5ex\hbox{\gtsima}} 
\def\lsim{\lower.5ex\hbox{\ltsima}} 
\def\gtsima{$\; \buildrel > \over \sim \;$} 
\def\ltsima{$\; \buildrel < \over \sim \;$} \def\gsim{\lower.5ex\hbox{\gtsima}} 
\def\lsim{\lower.5ex\hbox{\ltsima}} 
\def\simgt{\lower.5ex\hbox{\gtsima}} 
\def\simlt{\lower.5ex\hbox{\ltsima}} 
\newcommand{\ped}[1]{_{\mathrm{#1}}}

\def\cc{\rm cm^{-3}}

\def\S*{$\Sigma_{\rm SFR}$}

\def\kms{{\rm km\,s}^{-1}\,}

\definecolor{apcolor}{HTML}{b3003b}
\definecolor{afcolor}{HTML}{800080}
\definecolor{lvcolor}{HTML}{DF7401}
\definecolor{mdcolor}{HTML}{01abdf} 
\definecolor{cbcolor}{HTML}{ff0000}
\definecolor{sccolor}{HTML}{cc5500} 
\definecolor{sgcolor}{HTML}{00cc7a}

\makeatletter
\def\@hex@@Hex#1%
 {\if a#1A\else \if b#1B\else \if c#1C\else \if d#1D\else
  \if e#1E\else \if f#1F\else #1\fi\fi\fi\fi\fi\fi \@hex@Hex}
\makeatother

\newcommand\ozsobs{\Omega_{Z,*}^{\rm obs}}
\newcommand\odobs{\Omega_{d}^{\rm obs}}

\received{\today}
\revised{\today}
\accepted{\today}
\submitjournal{MNRAS}

\shorttitle{Cosmic dust evolution}
\shortauthors{Ferrara \& Peroux}
\graphicspath{{./}{figures/}}

\begin{document}

\title{Late-time cosmic evolution of dust: solving the puzzle}

\correspondingauthor{Andrea Ferrara}
\email{andrea.ferrara@sns.it}

\author[0000-0002-9400-7312]{Andrea Ferrara}
\affil{Scuola Normale Superiore,  Piazza dei Cavalieri 7, 50126 Pisa, Italy}

\author[0000-0002-4288-599X]{Celine Peroux}
\affiliation{European Southern Observatory,
Karl-Schwarzschild-Stra{\ss}e 2, 85748 Garching bei M\"unchen, Germany and\\
Aix Marseille Univ, CNRS, LAM, (Laboratoire d'Astrophysique de Marseille), UMR 7326, 13388, Marseille, France}

\begin{abstract}
Dust is an essential ingredient of galaxies, determining the physical and chemical conditions in the
interstellar medium. Several complementary observational evidences indicate that {the} cosmic dust mass density significantly drops from redshift $z=1$ to $z=0$. Clearly, and for the first time during cosmic evolution, dust must be destroyed more rapidly than it is formed. By considering the dust production/destruction processes acting in this cosmic time lapse, we find that the drop can be explained if dust is mainly destroyed by astration (49\% contribution in the fiducial case) and supernova shocks within galaxies (42\%). Our results further imply that on average each supernova destroys only $M_{d,sn} =0.45\, M_\odot$ of dust, {i.e. $5-10$ times less} than usually assumed, with a hard upper limit of $M_{d,sn} < {3.0} M_\odot$ set by the available metal budget and maximal grain growth. The lower efficiency might be explained by effective shielding of dust against shock processing in pre-supernova wind shells. 
\end{abstract}

\keywords{cosmology: observations -- intergalactic medium -- ISM: evolution -- dust, extinction}

\section{Introduction} \label{sec:intro}
Dust plays a crucial role in the thermal balance, dynamics and visibility of galaxies throughout cosmic times. Importantly, dust has a strong influence on the physical processes of the insterstellar medium (ISM) of galaxies in several ways. 

Grain surfaces and the Polycyclic Aromatic Hydrocarbons (PAH) participate in a large number of chemical reaction networks in different phases of ISM, and act as catalyst for important chemical processes such as the formation of H$\ped{2}$ \citep{Tielens10}, which in turn drives molecular chemistry. 

Dust governs the ISM thermal balance \citep{Draine03, Galliano18} by providing photoelectric heating, and cooling which can alter the shape of the Initial Mass Function (IMF) 
by favouring cloud fragmentation, thus inhibiting the formation of massive stars and fostering the formation of low-mass stars \citep{Schneider03,Omukai05}. 

Finally, grains absorb the stellar ultraviolet light and re-radiate it in the infrared, shielding the dense gas, and by these means triggering the formation of molecular clouds where new stars are born.  

In spite of the almost 80-years history of dust studies, relatively little is known about the origin and build-up history of the solid 
component of the ISM. The na\"ive expectation is that cosmic dust abundance should be tied to the 
metal abundance. However, recent data (presented in Sec. \ref{sec:data}) suggest that this is not the case. Indeed in the last $7-8$ Gyr the dust abundance has significantly decreased \citep[see Fig. 12 of][]{peroux2020} despite of the increasing availability of heavy elements, the primary components of dust grains.

Theoretically, a few studies have addressed the dust evolution issue. \citet[][see also \citet{Hou19}]{Aoyama18}, by performing cosmological simulations including dust evolution, found that the cosmic dust density\footnote{The cosmic evolution of the cosmic dust mass density, $\Omega_{d}$, is defined as the comoving density of dust in the Universe normalised by the critical density at redshift zero, $\Omega_{d} \equiv \rho_{d} / \rho_{\rm cr}(z=0)$.}, $\Omega_d(z)$, peaks at $z \approx 1$. They suggest that the slight decline afterwards is due to astration. A similar type of simulation has been presented by \citet{Li19}, who found that the total $\Omega_d$ (i.e. dust in galaxies and outside them) always increases with time; however, the comoving dust mass density excluding dust ejected out of galaxies via galactic winds peaks at  $z = 2$ and then declines.  They interpret this trend as a result of the reduced availability of gas-phase metals to be accreted on grains due to the decreasing star formation rate at $z \simlt 2$. In another study based on EAGLE \citep{Schaye15} simulations which assesses the reliability of SED fitting to recover the input dust mass, \citet{Baes20} derived an evolution of $\Omega_d(z)$ which fits well the flat trend measured by \citet{driver2018} data, but less so the steeper decline originally found by \citet{dunne2011}. 

In any case, the physical nature of the decline cannot be addressed by current hydrodynamical cosmological simulations which lack the detailed treatment of the dust formation/destruction processes, but instead simply scale the dust content with the metal abundance. While other numerical studies \citep[e.g.][]{Bekki15,McKinnon18, McKinnon19, Aoyama20, Osman20} have included some of these processes, these simulations concentrate on single, isolated galaxies, thus hampering the ability to use such important results in a cosmological framework. 

The reported decrease is not predicted by some semi-analytical models \citep[e.g.][]{Popping17}, but resembles that obtained by \citet{Gioannini17}, although their model does not account for dust destruction in the hot intracluster (ICM) and intragroup (IGrM) medium \citep[see also][]{Vijayan19, Triani20}.  

Notwithstanding these many modelling efforts, the current lack of convergence in the predictions indicates that the 
important issue of dust cosmic evolution is still, at best, poorly understood. 

{Our work is motivated by a single important question: why does the dust abundance -- for the first time during cosmic evolution -- decrease from $z=1$ to $z=0$ in spite of the increased availability of metals? We answer this question by combining new/recent observational data with simple but solid physical arguments. As a byproduct, we set novel constraints on dust destruction efficiency. The strength of the method is based on its simplicity. It complements more general models which need to make a larger number of assumptions and/or do not fully include dust physics.}

\section{Cosmic Dust Density: Observations}\label{sec:data}
The last decade has brought a wealth of new measurements of $\odobs$ based on different techniques, which together draw a coherent picture of the global evolution of the dust mass with cosmic times.
%
%
\begin{figure}
\centering\includegraphics[scale=0.6]{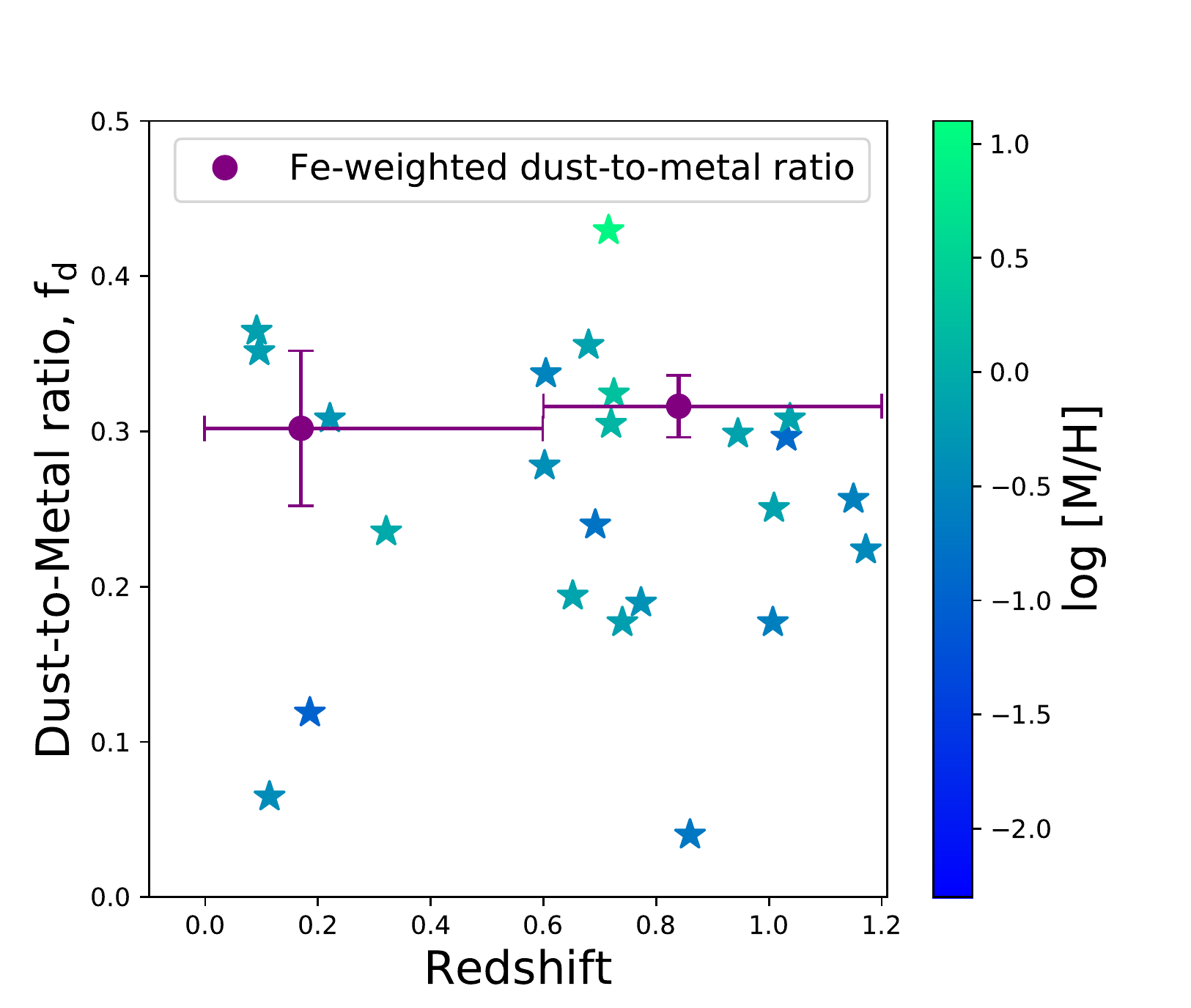}
\caption{Observations of dust-to-metal ratio in neutral gas traced by Damped Lyman-$\alpha$ Absorbers at $z<$1.2 \citep{peroux2020}. The individual measurements are shown by stars color-coded with metallicity, $\log$[M/H]. The purple error bars indicate the Fe-weighted mean dust-to-metal ratio, $f_{\rm d}$, for two redshift bins: $0<z<0.6$ and $0.6<z<1.2$. }
\label{Fig:DustToMetal}
\end{figure}
Initially, measurements came from infrared (IR) Spectral Energy Distribution (SED) fits of the extinction of individual galaxies. {We first describe techniques measuring the amount of dust in galaxies.} Making educated assumptions on the slope of the opacity power-law and the dust temperature, the IR emission of galaxies has been widely used to estimate their dust mass. The modelling of the IR SED has been especially improved in the past few decades with the arrival of far-IR ({\it Spitzer, Herschel}), submillimetre ({\it SCUBA, BLAST}), and {\it ALMA} ground instrumentation, adding much better constraints on the cold dust regime. 

In a work of reference, \cite{dunne2011} performed a measurement of the evolution of the dust mass density from a large sample of galaxies detected both at 250 $\mu$m in Herschel-ATLAS and in the Sloan survey. They reproduced the SED from temperature-based models fitted on the photometric data points. This work has been later complemented with large spectroscopic samples (including {\it GAMA}) and advanced SED fitting processes \citep[see also][]{clemens2013, clark2015, beeston2018,
driver2018, bellstedt2020}. Here, we use the value of {$\odobs$=$1.0^{+0.8}_{-0.5}\times 10^{-6}$ at $z=0$} derived from \cite{driver2018}, who despite a poorly constrained dust temperature, have a large sample leading to high statistical significance. {We note that our results are largely unsensitive to this choice though this is a conservative approach, as the value of $\odobs$ at $z=0$ derived by \cite{dunne2011} is even lower. }

Recently, \cite{pozzi2019} derived the evolution of the dust mass density from a far-IR (160$\mu$m) \textit{Herschel} selected catalogue in the COSMOS field, pushing estimates of $\odobs$ to $z=2.5$. They also find a broad peak at $z=1$, with a $\Omega_d$ decrease by a factor of $\approx 3$ from $z=1 \to 0$. Alternatively, ALMA deep fields allow to stack the contribution from, e.g., H-band selected galaxies and use the continuum detection at 1.2mm to derive the averaged dust mass in redshift bins over large lookback times \citep{magnelli2020}.

%
%
\begin{table*}[h]
\caption{Observed and expected cosmic dust and metal budget at late cosmic times. For the observed quantities, the redshift intervals are driven by the availability of data. The observed gas and dust values refer to neutral gas. They are recomputed refering to \cite{peroux2020} figures as follows: log [M/H] (their Fig. 7), $\Omega_{\rm gas}^{\rm obs}$ (their Fig. 3), $f_{\rm d}^{\rm obs}$ (their Fig. 10 \& 11; see eqn 1) and finally $\odobs$ (their Fig. 12). The metallicity values for stars ($\ozsobs$) and hot halos ($\Omega_{Z,h}^{\rm obs}$) are directly taken from \cite{peroux2020} (their Fig. 8). The expected quantities are derived from the following equations in the present work: $\Omega_*$ (eqn 4), $\Omega_Z$ (eqn 5) and $\Omega_d$ (eqn 6). The last column refers to the difference between $z\sim$0 and $z\sim$1 values defined as: $\Delta\Omega_j = \Omega_j (z\sim 0) - \Omega_j (z\sim 1)$. A positive $\Delta\Omega_j$ indicates an increase of the quantity with cosmic time. 
}

\label{tab:1}
\begin{center}
\begin{tabular}{lccc}
\hline \hline
                             & $z\sim$0                         &$z\sim$1                             &$\Delta\Omega_j$ \\
\hline
\textbf{Observed}            &                                  &                                     & \\
\hline
$z_{\rm mean}$               &$ 0.17^{+0.43}_{-0.17}$           &$ 0.84^{+0.36}_{-0.24}$              & \\
$\log [M/H]$                 &$-0.26^{+0.14}_{-0.11}$           &$-0.20^{+0.09}_{-0.09}$              & \\
$\Omega_{\rm gas}^{\rm obs}$ &$5.0^{+0.2}_{-0.3}\times 10^{-4}$ &$6.4^{+0.5}_{-0.4}\times 10^{-4}$   & \\
$f_{\rm d}^{\rm obs}$        &$0.30^{+0.05}_{-0.05}$            &$0.32^{+0.02}_{-0.02}$               & \\
$\odobs$                     &$1.0^{+0.8}_{-0.5}\times 10^{-6}$ &$1.6^{+0.3}_{-0.2}\times 10^{-6}$   &$-0.6^{+0.8}_{-0.5}\times 10^{-6}$\\
\hline
$z_{\rm mean}$               &$0.10^{+0.12}_{-0.09}$            & $0.70^{+0.05}_{-0.05}$              & \\
$\ozsobs$               &$3.39^{+1.18}_{-0.82}\times 10^{-5}$   &$0.91^{+0.95}_{-0.28}\times 10^{-5}$ & $+2.48^{+1.51}_{-0.87}\times 10^{-5}$ \\
$\Omega_{Z,h}^{\rm obs}$     &$10.0^{+3.5}_{-2.6}\times 10^{-6}$&$5.2^{+2.2}_{-1.5}\times 10^{-6}$    & $+4.8^{+4.1}_{-3.0}\times 10^{-6}$\\
\hline
\textbf{Expected}&&\\
\hline
$z_{\rm mean}$               & 0.0                              & 1.0                                 & \\
$\Omega_*$                   &$3.76\times10^{-3}$               &$2.33\times10^{-3}$                  &$+1.43\times 10^{-3}$\\
$\Omega_Z$                   &$1.24\times10^{-4}$               &$0.77\times10^{-4}$                  &$+0.47\times 10^{-4}$\\
$\Omega_d$                   &                                  &                                     &\hspace{-1.cm}$+(0.43,4.72)\times 10^{-5}$\\
\hline
\hline
\end{tabular}
\end{center}
\end{table*}

{An alternative technique aims at estimating the total dust mass in the Universe. To this end}, a number of Herschel surveys has been utilized to measure the far-infrared background anisotropy which captures the full population of grains responsible for thermal dust emission in galaxies. These measurements then provide an estimate of the global quantity of dust in the Universe \citep{de-bernardis2012, thacker2013}. These observations recover remarkably well the global evolution of $\odobs$ with cosmic times traced by individual galaxies. 

Lastly, a powerful approach to studying the dust content of the Universe is provided by {cold gas traced by} quasar absorbers. Indeed, the dust content of intervening gas has been assessed from the analysis of unrelated background objects. \cite{menard2010} derived an estimate for $\odobs$ using the reddening of SDSS quasars due to foreground Mg\,II absorbers, extending such measurements to higher redshift \citep[see also][]{menard2012}. This estimate  includes strong Mg\,II absorbers (with equivalent widths EW $>$0.8 \AA), which also trace the circum-galactic medium (CGM) of galaxies. \citet{peek2015} provided further estimates of the dust content of the CGM by integrating the stellar mass over the galaxy stellar mass function of \citet{wright2018}.  {These measurements provide an assessment of the dust content of galaxies' halos and therefore is complementary to SED-fitting techniques described above.}

In all of these cases, the dust mass is derived assuming an extinction curve and scaling the results based on the dust-to-gas ratio within the Small Magellanic Cloud. Adopting Milky Way dust properties increases the masses by a factor of $\approx1.8$. We note that \cite{Zafar2011, zafar2013b}, \cite{DeCia2013} and \cite{wiseman2017} show that depletion-based $A_V$ estimates differ from
reddening-based (i.e. SED fit) and postulate that extinction and reddening do not trace the same type of dust. \cite{wiseman2017} hint that
 reddening-based measurements might differ, since reddening is measuring the cumulative effect along the line-of-sight \citep{menard2012}. 
 
Until recently, models describing the chemical evolution in galaxies rely on dust studies in local metal-poor dwarfs \citep[e.g.][]{remy2014,de-vis2019}, which are usually adopted as a benchmark case for the interstellar medium in
galaxies at high-redshift \citep[see also][]{Shapley20}. An alternative approach was proposed by \cite{Jenkins09, DeCia16, Jenkins17, DeCia18a, DeCia18b, RomanDuval2019} who utilised multi-element methods to correct elemental depletion to estimate the amount of the metals locked into dust grains in neutral gas. These techniques enable to derive the dust-to-metal ratio ($f_d$) in quasar absorbers, extending such measurements to lower metallicities than are currently available in the local Universe and to higher redshifts than possible before. By combining these estimates with measurements of $\Omega_{\rm gas}^{\rm obs}$, \cite{peroux2020} uniquely derive the global dust density of the neutral gas. 
The individual dust-to-metal ratio measurements are displayed in Fig. \ref{Fig:DustToMetal} for $z<1.2$. We compute the Fe-weighted mean dust-to-metal ratios as follows:  
\begin{equation}\label{eq:obs}
    \langle f^{obs}_d \rangle = \frac{\Sigma(f_d\ \times N({\rm Fe}))}{\Sigma N({\rm Fe})},
\end{equation}
where the errors are estimated from the standard deviation,
$\sigma^{\prime}$:
\begin{equation}
	 \sigma^{\prime 2} = 
        \left(\sum
        ( f_d - 
         \langle f_d^{\rm obs}\rangle ) ^2\right)/(n - 1).
\end{equation}

{Therefore, the uncertainties do not take into account errors on the slopes of depletions versus [Zn/Fe]  \citep{DeCia16}, and on the contribution of carbon, an important contributor to dust mass. An additional uncertainty not considered here, is related to differential carbon depletion with respect to other elements \citep{Jenkins09}.}

Eq. \ref{eq:obs} represents the metal-weighted mean of the points binned by redshift interval. The results indicate that $f^{obs}_d$ in the cold phase remains constant over that redshift range. These values are shown in Fig. \ref{Fig:DustToMetal} and tabulated in Table~\ref{tab:1}. 
Taken together, this large set of observations depict a consistent picture
\citep[see Fig. 12 of][]{peroux2020}. Notwithstanding a continuous rise in the metal content of the Universe, the data indicate a surprising global decrease of the dust mass density from $z=1 \to 0$, of the order $\Delta \odobs = -0.6\times 10^{-6}$. This trend is readily apparent in all the results described above despite the many observational methods utilised. {To investigate this issue}, we next turn into making quantitative predictions of the expected amount of dust at late cosmic times.

\section{Expected Dust Density}\label{sec:exp}
The basic ingredient of the calculation is represented by the cosmic star formation history (CSFH), $\psi(z)$. Adopt the analytical fit to the available data provided by \citet{Madau14}:
\begin{equation}
    \psi(z) = \psi_0 \frac{(1+z)^{2.7}}{1+ [(1+z)/2.9]^{5.6}} \quad M_\odot {\rm yr}^{-1} {\rm Mpc}^{-3}, 
\end{equation}
with $\psi_0=0.015$. We can then compute the stellar mass density at any given redshift, $z$, by integrating:
\begin{equation}\label{rho*}
    \rho_*(z) = (1-R) \int_z^\infty \psi(z') \frac{dz'}{H(z')(1+z')} \quad M_\odot {\rm Mpc}^{-3}. 
\end{equation}
In the previous expression $H(z)=H_0 [\Omega_m(1+z)^3+\Omega_{\Lambda}]^{1/2}$ is the Hubble parameter; for consistency with data, we will adopt the following values of the cosmological parameters $(\Omega_m, \Omega_{\Lambda}, h) = (0.3, 0.7, 0.7)$. The return fraction of gas from stars appropriate for a Chabrier initial mass function and Instantaneous Recycling Approximation is $R=0.41$. The previous expression can be cast in a more handy form,
\begin{equation}
    \rho_*(z) = (1-R)\frac{\psi_0}{H_0} {\cal I}(z) = 1.26\times 10^8 {\cal I}(z) \quad M_\odot {\rm Mpc}^{-3}, 
\end{equation}
where the ${\cal I}(z)$ is the nondimensional integral in eq. \ref{rho*};
further define ${\cal I}^{x}\equiv {\cal I}(z=x)$, and note that $({\cal I}^{1},{\cal I}^{0})=(2.33, 3.76)$, and $\Delta {\cal I} \equiv {\cal I}^{0}-{\cal I}^{1} = 1.43$. For the adopted cosmology, the stellar density parameter is $\Omega_*=\rho_*/(3H_0^2/8\pi G) = 10^{-3} {\cal I}(z)$. The density of stars formed from $z=1$ to the present is $\Delta\Omega_* = 1.43\times 10^{-3}$. 

The metal density associated with $\Omega_*$ is
\begin{equation}
    \Omega_Z = y\, \Omega_* = 3.3 \times 10^{-5} {\cal I}(z); 
\end{equation}
we have assumed a metal yield $y=0.033 \pm 0.01$, with errors accounting for uncertainties in the nucleosynthetic yields \citep{Peeples14, Vincenzo16}. 

Finally, the dust mass can be computed by knowing the dust-to-metal ratio, i.e. the fraction of metals locked into dust, $f_d={\cal D}/Z$, where ${\cal D}$ and $Z$ are the dust-to-gas ratio and gas metallicity, respectively. In the Milky Way, ${\cal D}=1/162$ \citep{Zubko04}; assuming solar metallicity, $Z=0.0142$ \citep{Asplund09}, then $f_d=0.43$ \citep{Draine11}. At higher redshifts $f_d$ in measured from the gas depletion patterns in, e.g. Damped Lyman-$\alpha$ (DLA) systems \citep{peroux2020}. In the redshift range $0< z <1$, $f_d^{\rm obs} \approx 0.31$ (see Figure~\ref{Fig:DustToMetal} and Table \ref{tab:1}) in the cold, neutral medium out of which stars form. 

We note that $f_d^{\rm obs}$ combines the effects of dust production, growth and destruction, while we aim here at isolating the first process. Hence, we write the dust density associated with the metal density as 
\begin{equation}\label{omd_temp}
    \Omega_d = f_d \Omega_Z, 
\end{equation}
where, in general\footnote{{We stress that $f_d^{\rm obs}$ is refers to \textit{neutral gas} only, while $f_d$ and $\Omega_d$ denote the total dust abundance in all phases.}}, $f_d \neq f_d^{\rm obs}$. To determine $f_d$ we proceed as follows.

The minimum value, $f_d^{\rm min}$, is obtained by neglecting growth after grains are injected in the ISM by sources (SNe and AGB stars). SNe with progenitor mass $12-40\, M_\odot$ form on average $0.3 M_\odot$ of dust, 80\% of which is typically destroyed by the reverse shock on site \citep{Todini01, Bianchi07, Lesniewska19}. The AGB stars ($> 2 M_\odot$) contribution {per unit stellar mass formed} for $Z > 0.2-0.3 Z_\odot$ is $\approx 4$ times higher {than the SN one} \citep{Zhukovska08, DellAgli17, Valiante17}. Hence, the combined effective dust yield per supernova is $y_d=0.3 M_\odot$, which gives $f_d^{\rm min} = \nu y_d/y = 0.09$, where $\nu^{-1}= 102\, M_\odot$ is the number of SNe produced per stellar mass formed, according to the adopted \citet[eq. 17]{Chabrier03} IMF. 

Larger values of $f_d$ might arise as a result of grain growth, which depends on ambient conditions \citep{Asano13, Ferrara16}, and it is very difficult to estimate reliably. The maximum $f_d^{\rm max}=1$ value is obtained when all the metals are depleted (maximal growth efficiency). Note that typical values observed in galaxies fall conveniently in the range $f_d^{\rm min} < f_d^{\rm obs} < f_d^{\rm max}$.  To account for dust growth uncertainties we write the dust density as            
\begin{equation}\label{omd}
    \Omega_d = (f_d^{\rm min}, f_d^{\rm max}) \Omega_Z = (0.3,3.3) \times 10^{-5} {\cal I}(z). 
\end{equation}

\begin{figure*}
\includegraphics[scale=0.5]{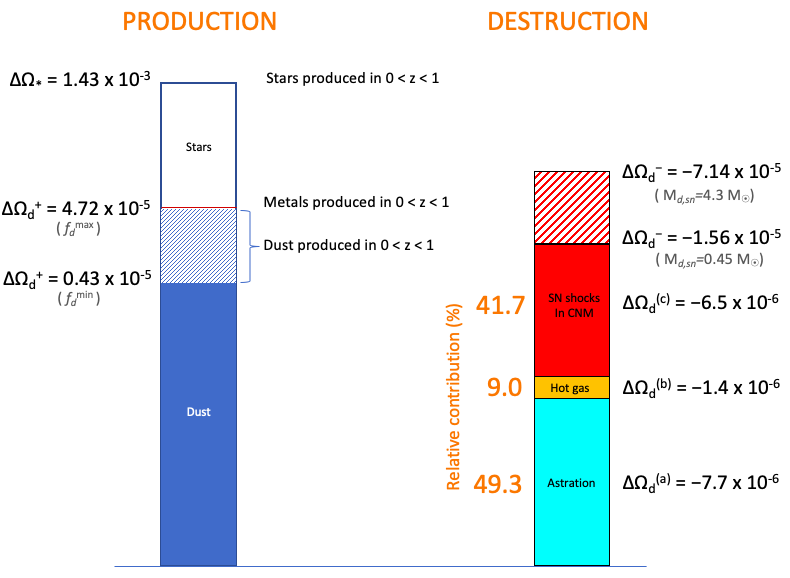}
\centering\caption{Sketch of dust production and destruction mechanisms and corresponding yields described in Sec. \ref{sec:model}. The diagonally-hatched blue region in the production histogram denotes the uncertainty on the $f_d$ value of dust sources; {errors on each individual $\Delta\Omega_d^{(i)}$ contribution are given in Tab. \ref{tab:1} and in Sec. 4.}. Relative contributions to the destruction budget refer to the fiducial, low SN destruction efficiency case ($M_{d,sn}=0.45 M_\odot$).}
\label{Fig02}
\end{figure*}

Fig. \ref{Fig02} {shows} the detailed balance of dust production and destruction (discussed in the next Sec.) mechanisms, and their relative importance.
As ${\cal I}(z)$ is a {decreasing} function of redshift, the expected cosmic dust content should steadily increase with time, reaching $\Omega_d= (1.13,12.41)\times 10^{-5}$ at $z=0$, with a variation $\Delta\Omega_d^+ = +(0.43, 4.72) \times 10^{-5}$ from $z=1$ in the case of (zero, maximal) dust growth efficiency. 

This conclusion is in striking contrast with observations, which indicate a decrease of $\Omega_d$ from $z=1 \to 0$ ($\Delta \Omega_d^{\rm obs}=-0.6\times 10^{-6}$, see Table \ref{tab:1}). Clearly, and for the first time during cosmic evolution, dust must be destroyed more rapidly than it is formed during this time span of 7.8 Gyr \citep{gjergo2020}.

\section{Dust destruction} \label{sec:model}
We now consider the various dust destruction processes at play. Before we proceed, we justify the assumption we will make that $f_d=0$ in hot ($T\simgt 10^5$ K) gas, such as the ICM/IGrM, or in SN-driven galactic outflows. In these environments dust is destroyed by thermal sputtering with ions and electrons of the plasma. The rate at which the grain radius decreases is described by a simple fit to the numerical results by \citet{Draine79b, Tsai95, Dwek96}:
\begin{equation}\label{sput} 
\frac{da} {dt} = - A n T_6^{-1/4} e^{-BT_6^{-1/2}},
\end{equation}
where $a$ is the grain size, $n$ and $T_6=T/10^6\,{\rm K}$ are the gas density and temperature; we have adopted material-averaged values for the constants $(A, B)=(1.2\times 10^{-5} \mu$m yr$^{-1}, 3.85)$. From eq. \ref{sput} the survival time of a typical grain ($a=0.1\, \mu$m) in a $T=10^6$ K gas is $\tau_s = (0.4/n)$ Myr. Provided $n > 5\times 10^{-5}\, \cc$, which applies to ICM/IGrM, grains produced at $z=1$ are destroyed by thermal sputtering well before $z=0$. While the details of the process depend on the exact gas temperature, grain size distribution, and residence time, assuming $f_d=0$ in the hot cosmic gas appears warranted.    

Let us go back to the three main dust destruction processes and quantify their impact. These are: (a) astration, (b) thermal sputtering, (c) supernova (SN) shocks; they are discussed separately in the following. Their combined action must lead to a decrease the dust mass density at $z=0$ to the observed value $\Omega_d^{\rm obs} = 10^{-6}$. 

%
%

\subsection{Destruction by astration}
Astration involves the incorporation of gas and dust into a stellar interior during star formation. As stars forms in cold, neutral gas, we assume that the stellar build-up material has $f_d = f_d^{\rm obs}=0.31$. 
The data in Table \ref{tab:1} show that the increase of metals in stars is $\Delta\Omega_{Z,*}^{\rm obs}= 2.48\times 10^{-5}$. Then, the (negative) variation of dust density due to astration is
\begin{equation}
    \Delta\Omega_d^{(a)} = - f_d^{\rm obs} \Delta\Omega_{Z,*}^{\rm obs} = - 7.7{^{+4.9}_{-3.0}}  \times 10^{-6}.
\end{equation}
Astration contributes ($11-49$)\% (for a high or low SN dust destruction efficiency, respectively; see Sec. \ref{dest}) to the total amount of dust destruction; the rest must be removed by the other two mechanisms.

\subsection{Destruction by hot gas}
As already mentioned, we assume that as dust gets embedded in the hot phase, it is -- for our purposes -- instantaneously and completely eroded by sputtering. This implies that dust associated with metals contained in the hot cosmic gas at $z=0$ must be removed from the total budget. The metal content of hot gas has increased from $z=1 \to 0$ by $\Delta\Omega_{Z,h}^{\rm obs}= 4.8\times 10^{-6}$ (see Table~\ref{tab:1}). This term contributes a negative variation equal to  
\begin{equation}
    \Delta\Omega_d^{(b)} = - f_d^{\rm obs} \Delta\Omega_{Z,h}^{\rm obs} = - 1.4{^{+1.3}_{-1.0}}  \times 10^{-6},
\end{equation}
corresponding to a mere ($2-9$)\% of the total dust destruction budget. 

\subsection{Destruction by SN shocks in galaxies}
Finally, we consider dust destruction by SN shocks in the ISM of galaxies, which we identify here with the cold, neutral gas. Recent detailed numerical simulations of dust production and destruction in SN explosions \citep{Martinez19} find that, once the presence of a pre-supernova wind-driven cavity is properly included, dust destruction is strongly suppressed. The physical reason for this is that the dust is collected in a dense shell by the wind; the shell represents an almost insurmountable barrier that prevents the SN blast wave from processing the majority of the ambient dust protected by the shell. As a result, under typical ambient conditions (gas density\footnote{Cases with  $n=10^3\, \cc$ have also been explored, showing a $\sim25$\% decrease in the amount of dust destroyed.} $n \approx 1\, \cc$), the amount of dust destroyed\footnote{\citet{Hu19} performed similar simulations finding $M_{d,sn} \approx 5 M_\odot$ for $n = 1\, \cc$. They do not treat the pre-supernova wind-driven cavity self-consistently, but when they allow SNe to occur in hot ($T>10^4$ K) bubbles carved by previous SN explosions, the destruction rate is decreased by a factor $\approx 2.5$.  We point out that reduced destruction due to the pre-SN wind affects \textit{each} SN, not only those exploding in pre-existing hot bubbles.} per SN event is $M_{d,sn} = 0.45 M_\odot$. 

An alternative calculation, which however does not include the effects of the wind-driven shell discussed above, might be performed as follows. First, dust sputtering requires projectiles (electrons, ions) with kinetic energies $E_t> 100$ eV \citep{Draine79a}, which can be produced by shocks with velocities $v_s \simgt 200\, \kms$. As the transition from the energy-conserving, Sedov-Taylor phase to the radiative one occurs at $v_s = 200 (n^2 E_{51})^{1/14}\, \kms$ \citep{McKee89}, we conclude that dust destruction essentially terminates with the first phase, unless the density is very high (typically, though, the diffuse gas component in galaxies has the largest filling factor; hence, expansion primarily occurs in low density gas). The mass swept-up by the shock, $M_e$, as a function of $v_s$ is
\begin{equation}\label{Me}
    M_{e}(v_s) = \frac{E}{\sigma v_s^2} = 6800 \frac{E_{51}}{v_{s7}^2} \quad M_\odot;
\end{equation}
where $\sigma=0.736$, $E_{51}= E/10^{51} \rm erg\, {= 1}$ is the explosion energy, and $v_{s7}= v_s/100\,\kms$. Hence, $M_e(v_s)=1700\, M_\odot$ {for $v_s=200\, \kms$}. The dust destruction efficiency, $\gamma_d(v_s)$, by a shock depends on its velocity. For $n=0.25\, \cc$,  \cite{Slavin15} give the following fit to their numerical results, valid for $1.85 < v_{s7}< 5$: 
\begin{equation}\label{eff}
    \gamma_d = -1.9+2.02v_{s7}-0.641v_{s7}^2+0.092v_{s7}^3-0.05 v_{s7}^4
\end{equation}
By mass-averaging $\gamma_d$ using expression eq. \ref{Me}, we find $\bar \gamma_d= 0.41$. Hence, the dust mass destroyed per SN in this case, assuming solar metallicity gas, would be $M_{d,sn} = \bar \gamma_d\, {\cal D}\, M_e(v_t) = 4.3\, M_\odot$, i.e. a factor about $10$ times higher than obtained by \citet{Martinez19}. Given these uncertainties we will use these values to bracket our results. 

The number density of SN exploded in $0< z <1$ is  $\Delta{\cal N} = \nu \Delta \rho_* = 1.26\times 10^8 \nu \Delta {\cal I} = 1.8 \times 10^6\, {\rm Mpc}^{-3}$, or 
\begin{equation}\label{Mswept}
    \Delta\Omega_d^{(c)} = - M_{d,sn} \Delta{\cal N}\rho_{cr}^{-1} = - (6.5-62.3)  \times 10^{-6},
\end{equation}
depending on the adopted value of $M_{d,sn}$. 

\subsection{Total dust destruction}\label{dest}
Figure~\ref{Fig02} displays the results in graphic form. In summary, processes (a)--(c) account for a total dust destruction corresponding to 
\begin{equation}\label{od-}
\Delta \Omega_d^{-}\equiv\Sigma_{i} \Delta\Omega_d^{(i)} = - (0.91{^{+0.5}_{-0.3}} +1.45\, M_{d,sn})\times 10^{-5}, 
\end{equation}
where $M_{d,sn}$ is in solar masses, and $i$=(a,b,c). This value must balance the dust mass produced in $0<z<1$, $\Delta\Omega_d^+$, augmented by the observed dust decrease during the same epoch, $\Delta\odobs=-0.6{^{+0.8}_{-0.5}}\times 10^{-6}$:
\begin{equation}\label{bal}
    \Delta \Omega_d^{+} - \Delta\odobs + \Delta \Omega_d^{-} = 0
\end{equation}
From eqs. \ref{od-} and \ref{bal} we can then conclude that: (a) if dust growth does not occur ($f_d=f_d^{\rm min}$), then {$\Delta \Omega_d^{-} \simgt \Delta\odobs -\Delta \Omega_d^{+}$ within errors, implying that observed $\Omega_d$ decrease can be explained by astration and sputtering in hot gas only, without the need for} dust destruction by SNe (i.e. $M_{d,sn}=0$); (b) the low-efficiency SN destruction ($M_{d,sn}=0.45 M_\odot$) would yield $\Delta \Omega_d^- = -1.56{^{+0.5}_{-0.3}}\times 10^{-5}$, which falls exactly in the middle of the allowed production/growth range. {Then, by combining eqs. \ref{bal} and \ref{omd}, we get a nominal value} $f_d=0.34$, a value tantalizingly close to the observed one (0.31); (c) assuming that all the newly produced metals in $0 < z < 1$ are incorporated into dust ($f_d=f_d^{\rm max}=1$), we can get a hard {upper limit on $M_{d,sn}< 3.0 M_\odot$}. Larger values, such as those ($4.3 M_\odot$) predicted by the high SN destruction efficiency case, would produce a steeper $\Omega_d$ decrease at late times, and are therefore inconsistent with the data.   

\section{Implications for dust physics} \label{sec:sol}
Most likely, the usually adopted destruction rates in SN shocks have been significantly overestimated as they result in an apparent inconsistency with the observed cosmic dust evolution. From our calculation, combined with the available data, we conclude that each SN can destroy at most ${3.0}\, M_\odot$ of dust. This upper limit assumes that all the metals are locked into dust; most likely, the actual value is a factor 4-5$\times$ lower, in agreement with recent theoretical findings \citep{Martinez19}.   

At the same time, \citet{Ferrara16} noted that dust growth, particularly at high-$z$, is problematic and invoked solutions in which a lower growth rate is balanced by a reduced destruction rate as we suggest here. Finally, we notice that our empirical argument, based on dust cosmic evolution, resonates with theoretical down-revaluations of the dust destruction rates by SN presented by \citet{Jones11}. 

\section{Summary}
We have investigated the evolution of the cosmic dust density in the last $\approx 8$ Gyr. During this time stretch (corresponding to $z= 1 \to 0$), observations show that $\Omega_d$ has \textit{decreased} by about $37.5\%$ in spite of the fact that the cosmic metal abundance has \textit{increased} by about a factor 1.6. 
Thus, dust must have been efficiently destroyed during this period. 

By evaluating different dust destruction mechanisms, we conclude that astration and SN shocks in the ISM of galaxies are the dominant factors, with sputtering in hot gas playing a sub-dominant role. All these processes were obviously at work also at $z>1$, but the decrease of $\Omega_d$ at later times is driven by the declining cosmic star formation rate and associated metal production.

An implication of our study is that the dust destruction efficiency required to explain the data is $\approx 10$ times lower than usually adopted (i.e. 0.45 $M_\odot$ vs. 4.3 $M_\odot$ of destroyed dust/SN) as suggested by recent hydrodynamical simulations \citep{Martinez19} leading to a reduced efficiency caused by the shielding effects of pre-SN wind-driven shells. By assuming a maximally efficient grain growth in the ISM, we find that the available metal budget sets a hard upper limit $M_{d,sn}<{3.0} M_\odot$ on the dust mass destroyed per SN.   

\acknowledgments
We thank V. D'Odorico, G. Popping, L. Sommovigo for insightful comments. AF acknowledges support from the ERC Advanced Grant INTERSTELLAR H2020/740120. Any dissemination of results must indicate that it reflects only the author’s view and that the Commission is not responsible for any use that may be made of the information it contains. Generous support from the Carl Friedrich von Siemens-Forschungspreis der Alexander von Humboldt-Stiftung Research Award is kindly acknowledged (AF). AF thanks the European Southern Observatory (ESO) and Max-Planck for Astrophysics (MPA) in Garching for a warm hospitality during part of this research. CP is grateful to the Alexander von Humboldt-Stiftung for the granting of a Bessel Research Award. CP is indebted to the Max-Planck for Astrophysics (MPA) in Garching for a fruitful visit.
All plots in this paper were built with the \textsc{matplotlib}  \citep{Hunter07} package for \textsc{PYTHON}.\\    
{\bf Data Availability}\\
Data available on request.

\bibliography{paper}{}
\bibliographystyle{aasjournal}



\end{document}